\documentclass[aps,twocolumn,showpacs]{revtex4}

\usepackage{graphicx}

\begin{document}

\title{Towards device-size atomistic models of amorphous silicon}

\author{R. L. C. Vink}
\email{vink@phys.uu.nl}
\homepage{http://www.phys.uu.nl/~vink}
\affiliation{
        Institute for Theoretical Physics,
        Utrecht University,
        Leuvenlaan 4,
        3584 CE Utrecht, the Netherlands }

\author{G. T. Barkema}
\affiliation{
        Institute for Theoretical Physics,
        Utrecht University,
        Leuvenlaan 4,
        3584 CE Utrecht, the Netherlands }

\author{M. A. Stijnman}
\affiliation{
	Mathematical Institute, 
	Utrecht University, 
	PO Box 80010, 
	3508 TA Utrecht, the Netherlands }

\author{R. H. Bisseling}
\email{Rob.Bisseling@math.uu.nl}
\homepage{http://www.math.uu.nl/people/bisseling}
\affiliation{
	Mathematical Institute, 
	Utrecht University, 
	PO Box 80010, 
	3508 TA Utrecht, the Netherlands }

\date{\today}

\begin{abstract} 
The atomic structure of amorphous materials is believed to be well
described by the continuous random network model. We present an algorithm
for the generation of large, high-quality continuous random networks. The
algorithm is a variation of the {\it sillium} approach introduced by
Wooten, Winer, and Weaire. By employing local relaxation techniques, local
atomic rearrangements can be tried that scale almost independently of
system size. This scaling property of the algorithm paves the way for the
generation of realistic device-size atomic networks.
\end{abstract}



\maketitle

\section{Introduction}

The structure of amorphous semiconductors is believed to be well
represented by the continuous random network (CRN) model introduced by
Zachariasen more than sixty years ago~\cite{zachariasen32}. As a result,
the generation of high quality CRNs has been the subject of investigation
for many years. The first CRNs were built by hand, see for instance the
work of Polk~\cite{polk}. Nowadays, the generation of CRNs is mostly
carried out on computers.

The first computer-generated networks, which date back to the sixties and
seventies, typically contain a few hundred particles. More advanced
algorithms and faster computers have increased the size of the networks
that can be handled to a few thousand atoms, with simulation cells of up
to \mbox{$40\times 40\times 40$ \AA$^3$}. As the simulation cells increase
in size, actual devices have decreased in size. For example, the thickness
of solar cells based on amorphous silicon has already decreased to
\mbox{1000 \AA}; and because in-plane periodicity after approximately
\mbox{30 \AA}\ is expected to be a good approximation of the macroscopic
lateral size, a reasonable solar cell model would require a simulation
cell of \mbox{ $30\times 30\times 1000$ \AA$^3$}, containing approximately
45,000 atoms. This is only one order of magnitude larger than currently
feasible.  For other electronic devices, lithography on \mbox{0.1 $\mu$m
(=1000 \AA)} technology is expected to be reached in the coming decade.

In this work, we present a computational approach to generating large
CRNs, and discuss the properties of high quality networks containing up to
20,000 particles. This achievement shows that the generation of
device-size atomic configuration networks is within reach.

We begin by describing the algorithm of Wooten, Winer, and Weaire (WWW),
which has been the basis of the best CRNs generated to date. We then move
on to describe a number of improvements made to the original WWW algorithm
by Barkema and Mousseau in 1999~\cite{barkemahq}. These improvements
accelerate the relaxation by two orders of magnitude or more. Both the
original and the improved WWW algorithm, however, scale poorly with system
size, since the computational effort per attempted local atomic
rearrangement increases linearly with system size. In this work, we
introduce local force and energy evaluations and improve the scaling of
computation time with system size $N$ significantly, namely to a constant
per attempted move plus $\cal O$$(N)$ per accepted move. We also
demonstrate how parallel processing can be used to realize an additional
speedup, with parallel efficiencies of over 50\%.  The significance of
these improvements is demonstrated by generating 10,000-atom and
20,000-atom CRNs. We then discuss the structural and electronic properties
of these models and conclude with an outlook on future research, aiming
towards the generation of device-size atomic networks.

\section{The WWW algorithm}

\begin{figure}
\begin{center}
\includegraphics[width=6cm]{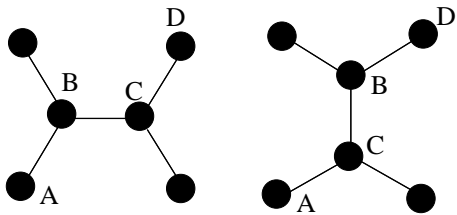}
\caption{\label{fig:wwwmove} Diagram depicting the WWW bond transposition.  
Four atoms A, B, C, and D are selected following the geometry shown left;
two bonds, AB and CD, are then broken and atoms A and D are reassigned
to C and B, respectively, creating two new bonds, AC and BD shown right.}
\end{center}
\end{figure}

In 1985, Wooten, Winer, and Weaire presented an algorithm for the
generation of four-fold coordinated CRNs~\cite{www}. In their approach, a
configuration consists of the coordinates of $N$ atoms and a list of the
$2N$ bonds between them. The structural evolution consists of a sequence
of bond transpositions as illustrated in Fig.~\ref{fig:wwwmove}.

Within the original WWW approach, the generation of a CRN starts with a
cubic diamond structure which is randomized by a large number of such bond
transpositions. After randomization, the network is relaxed through a
sequence of bond transpositions, accepted with the Metropolis acceptance
probability~\cite{metrop}:
\begin{equation}
\label{eq:metropolis}
  P = \min \left[1, \exp \left( \frac{E_b-E_f}{k_{\mathrm{B}} T} \right)  \right],
\end{equation}
where $k_{\mathrm{B}}$ is the Boltzmann constant,
$T$ is the temperature, and $E_b$ 
and $E_f$ are the total quenched energies of the system before and after the 
proposed bond transposition.

With an explicit list of neighbors, it is possible to use a simple
interaction such as the Keating potential~\cite{keating66} to calculate
energy and forces:
\begin{eqnarray}\label{eq:keating}
  E &=& \frac{3}{16} \frac{\alpha}{d^2} \sum_{<ij>} 
	\left( \vec{r}_{ij} \cdot \vec{r}_{ij} - d^2\right)^2 \nonumber \\
    &+& \frac{3}{8} \frac{\beta}{d^2} \sum_{<jik>} 
	\left( \vec{r}_{ij} \cdot \vec{r}_{ik} + \frac{1}{3} d^2 \right)^2,
\end{eqnarray}
where $\alpha$ and $\beta$ are the bond-stretching and bond-bending force
constants, and $d=2.35$ \AA\ is the Si-Si strain-free equilibrium bond
length in the diamond structure. Usual values are $\alpha=2.965$ eV
\AA$^{-2}$ and $\beta=0.285\alpha$.

With the approach described above Wooten and Weaire generated a 216-atom
model with an angular distribution as low as 10.9 degrees~\cite{ww}. A
decade later, using the same approach but more computing power,
Djordjevi\'c, Thorpe, and Wooten produced two large 4096-atom networks of
even better quality, with a bond-angle distribution of 11.02 degrees for
configurations without four-membered rings and 10.51 degrees when these
rings are allowed~\cite{djordjevic95}.

\section{The improved WWW algorithm}

The WWW algorithm in its original form is capable of producing high
quality amorphous networks containing of the order of a thousand atoms; it
is not well suited to generate much larger networks. This is mostly due to
the fact that for each proposed bond transposition, about one hundred
energy and force calculations are required, each scaling as $\cal O$$(N)$
with system size $N$. These $\cal O$$(N)$ operations are the bottleneck of
the algorithm.

In 1999, Barkema and Mousseau (BM) presented a number of modifications
to the original WWW algorithm, partially aimed at resolving these poor
scaling properties~\cite{barkemahq}. Their modifications are summarized
below:
\begin{enumerate}
\item Starting point for the relaxation in this case is a truly random
configuration whereby the atoms are placed at random locations in a
periodic box at the crystalline density. This guarantees that the
resulting network is not contaminated by some memory of the crystalline
state.
\item After a bond transposition in the original WWW approach, the
structure is always completely quenched, i.e., all the atomic coordinates
are fully relaxed. After the quench, the bond transposition is either
accepted or rejected based on the Metropolis probability. In contrast, BM
determines a threshold energy before quenching. During the quench the
final quenched energy is continuously estimated. Relaxation is stopped
when it becomes clear that the threshold energy cannot be reached so that
the bond transposition will eventually have to be rejected. This leads to
a large reduction in the number of force evaluations associated with
rejected bond transpositions.
\item A local relaxation procedure is used whenever possible. Immediately
after a bond transposition, only a small cluster of atoms in the model
experiences a significant force. This cluster consists of the atoms
directly involved in the bond transposition (marked A, B, C, and D in
Fig.~\ref{fig:wwwmove}) and of nearby atoms, typically up to the fourth
neighbor shell of the four transposition atoms. The number of atoms in
such a cluster is about 150. It therefore suffices to calculate the force
{\it locally} (i.e.,~only for the 150 or so atoms inside the cluster)
rather than {\it globally} (i.e., for all the atoms in the model).

Calculating the force on a cluster of atoms is an $\cal O$(1) operation,
which means that it is independent of the total system size. Local force
calculations are therefore much cheaper than global $\cal O$$(N)$ force
calculations. By using a local relaxation scheme we can increase the
efficiency of the algorithm significantly. Still, to make the final
accept/reject decision on the proposed move, the total Keating energy of
the system has to be calculated, which is again an $\cal O$$(N)$
operation. In practice, a switch must be made from local to
global relaxation, usually after about ten local relaxation steps.
\item The zero temperature case is treated specifically.
\end{enumerate}

Using the improved WWW algorithm, Barkema and Mousseau generated two
1000-atom models with bond angle deviations as low as 9.20
degrees~\cite{barkemahq}.  Furthermore, using the same algorithm they
generated a 4096-atom model with an angular deviation of 9.89 degrees.
All models show structural and electronic properties in excellent
agreement with experiments.

\section{A scalable WWW algorithm}

While the improved WWW algorithm can successfully generate networks
containing several thousand particles, it does not deal well with systems
of 10,000 particles or more. Each attempted bond transposition still
requires one or more $\cal O$$(N)$ energy evaluations. In this section, we
present an algorithm for attempting bond transpositions that is completely
local, i.e.,~free of $\cal O$$(N)$ operations for unsuccessful bond
transpositions.

\subsection{Local energy and force evaluations}

To exploit the local nature of the bond transpositions, we need to
introduce the concept of local energy: we assign to each atom $i$ an
energy $\epsilon_i$ such that \mbox{$E = \sum_{i=1}^N \epsilon_i$} with
$E$ the Keating energy of the system given by Eq.~(\ref{eq:keating}). One
way to achieve this is to divide the energy due to two-body interactions
equally between the two participating atoms and to assign the energy of
three-body interactions to the central atom of the corresponding triple.
Thus, we obtain:
\begin{eqnarray}\label{eq:localenergy}
  \epsilon_i &\equiv& \sum_{j=1}^4 \left[ \frac{1}{2} \frac{3}{16} \frac{\alpha}{d^2} 
	\left( \vec{r}_j \cdot \vec{r}_j - d^2\right)^2 \right. \nonumber \\ 
  &+& \left. \sum_{k=j+1}^4 \frac{3}{8} \frac{\beta}{d^2}
        \left( \vec{r}_j \cdot \vec{r}_k + \frac{1}{3} d^2 \right)^2  \right].
\end{eqnarray}
Here, the constants $\alpha$, $\beta$, and $d$ are defined as in
Eq.~(\ref{eq:keating});  $\vec{r}_j$ represents the vector pointing in the
direction of the $j$-th bond away from atom $i$. The energy $E_c$ of a 
cluster $C$ of atoms can now be calculated using:
\begin{equation}
\label{eq:clusterenergy}
  E_c = \sum_{i \in C} \epsilon_i.
\end{equation}

The force on the atoms inside the cluster is obtained from the derivative
of the Keating energy with respect to the atomic coordinates. Care has to
be taken for atoms on the edge of the cluster since these atoms also
interact with atoms outside the cluster: due to the two- and three-body
terms in the Keating potential, all atoms interact with their first and
second nearest neighbors; for atoms located on the edge of the cluster,
some of these neighbors are outside the cluster.

\subsection{Local WWW moves}

Starting point is a random configuration generated using the method
described in Ref.~\onlinecite{barkemahq}. This guarantees that the
resulting configurations are not contaminated by some memory of the
crystalline state. Assuming that the total Keating energy of the initial
configuration is known and equals $E$, WWW moves can be attempted locally
as follows:
\begin{enumerate}

\item A threshold energy $E_t$ is determined by using the equation:
\begin{equation}
  E_t = E - k_{\mathrm{B}} T \ln(1-r),
\end{equation}
where $r$ is a random number uniformly drawn from the interval $[ 0,1
\rangle$. The move is accepted if the attempted bond transposition leads
to a configurational energy below the threshold energy; otherwise it is
rejected.

\item The four atoms involved in the attempted bond transposition and all
atoms up to the fourth neighbor shells of these four atoms are grouped
into a cluster. Such a cluster contains about 150 atoms.

\item A list is constructed of all the bonds that contribute to the force
on the atoms inside the cluster. As was explained above, some of these
bonds involve atoms outside the cluster. For each bond we store the labels
of the two atoms constituting the bond, the $x$, $y$, and $z$ components
of the bond vector (taking care of the periodic boundary conditions), and
the square of the bond length. We then calculate the cluster energy $E_c$
using Eq.~(\ref{eq:clusterenergy}). In the calculation of the cluster
energy most bonds are encountered more than once. To increase efficiency,
a bond (i.e., its set of three components) is calculated only once during
an energy or force evaluation; once a bond has been calculated it is
time-stamped with an integer flag and the bond information is stored.  
Later references to the same bond are then retrieved from memory.

We also store the energy of the atoms that remain outside the cluster:
\mbox{$E_r \equiv E - E_c$}. We then perform the bond transposition to
obtain the geometry shown in the right frame of Fig.~\ref{fig:wwwmove}.

\item The system is relaxed locally, i.e., only atoms inside the cluster
are allowed to move. At each relaxation step we use
Eq.~(\ref{eq:clusterenergy}) to calculate the energy of the cluster $E_c$
and the atomic forces, again making sure each bond is calculated only
once, and perform structural relaxation as in the original and improved
WWW algorithms.  At each relaxation step the total energy of the system is
equal to \mbox{$E=E_r+E_c$}.  Local relaxation is continued until the
energy has converged or until it becomes clear that the threshold energy
cannot be reached.
\end{enumerate}

In the local relaxation procedure above, the computational effort per
attempted bond transposition does not grow with the system size.  Local
relaxation alone, however, is not sufficient and we also have to relax
globally to relieve any strain that may have built up between atoms on the
edge of a cluster and non-cluster atoms. For clusters extending up to the
fourth neighbor shell around the atoms directly involved in the bond
transposition we find that global relaxation can lower the configurational
energy typically by less than \mbox{0.1 eV}. We therefore switch from
local to global relaxation when, during local relaxation, the energy comes
to within \mbox{0.1 eV} of the threshold energy. In most cases, this leads
to the move being accepted.

\section{Parallel processing}

We have developed a parallel version of our algorithm with the aim of
harnessing the tremendous power of parallel computers. The parallel
algorithm is in bulk synchronous parallel (BSP) style~\cite{valiant90}
with alternating phases of computation and communication, separated by a
synchronization of all the processors. The parallel algorithm has two main
parts, local relaxation and global relaxation.

The local relaxation is done in parallel by letting every processor try a
sequence of randomly chosen bond transpositions, until one of the
processors finds an acceptable transposition. The processors work
independently but synchronize at regular intervals to communicate their
success or failure to the others. If more than one processor succeeds, an
arbitrary bond transposition is chosen as the winner. This approach
requires the replication of all the atomic data. Fortunately, the memory
storage needed is limited to an array of $3N$ atomic coordinates and a few
other arrays of size $N$, which usually can be stored on every processor.
(The WWW algorithm and its variants are demanding in CPU time, but not in
memory requirements.) Furthermore, this approach also requires refreshing
the atomic data when the positions change, causing communication between
the processors. This only happens after a bond transposition is accepted,
which is a relatively rare event (of the order of once every thousand
attempts). For these reasons, we choose to replicate the data instead of
distributing them, and develop a parallel local relaxation algorithm based
on replicated data.

It is crucial to choose a suitable time interval between successive
synchronizations. If this interval is too short, the time of the
synchronization itself will become dominant, or fluctuations in the amount
of work of the different processors will become visible; in longer
intervals such fluctuations are averaged out and have less impact. If the
interval is too long, it becomes likely that one (or more) accepted moves
are found in every time interval. Part of the work in a successful time
interval is wasted, because the processor that finds an accepted move
waits until the others have finished their (useless) computations. Thus a
high success rate means that much time is wasted. From the point of view
of parallel efficiency, the ideal situation occurs when most time
intervals fail to produce a successful bond transposition. On average,
little CPU time is then wasted.

A processor decides to synchronize based on the total number of relaxation
iterations performed during all its bond transposition attempts. Each
iteration requires of the order of $10^5$ floating point operations
(flops). Simply counting the number of attempts would not give a good
indication of the total amount of work performed by a processor, since the
number of iterations per attempt may vary, depending for instance on the
observed energy decrease. Thus, a processor synchronizes after every $b$
iterations. We determined the parameter $b$ empirically, and found on our
machine, a Cray T3E, that values in the range $b$=10--200 give the best
performance on $p=8$ processors; for higher numbers of processors this
range becomes smaller and the choice of $b$ becomes critical. For $p=32$,
we used $b=50$. The BSP cost model~\cite{valiant90} can be helpful in
choosing $b$. For instance, the BSP parameter $l$, representing the
synchronization time of the parallel computer, can be used to find a lower
bound for $b$.

The global relaxation is done in parallel by partitioning the simulation
cell over the $p$ processors of the parallel computer and letting every
processor compute the energies $\epsilon_i$, forces, and displacements for
the atoms in its own part of the cell. In contrast to the local
relaxation, it is now justified to have all the processors participate in
one relaxation: the amount of work, $\cal{O}$$(N)$, in an iteration is
much more than in the case of the local relaxation. In fact, processors
are even obliged to participate, because there is no other useful work to
do: most likely the global relaxation succeeds and provides the starting
point for the remainder of the computation.

Communication arises in the global relaxation because processors need data
from other processors concerning atoms that lie near inter-processor
boundaries. Thus at the end of an iteration, a processor has to
communicate the changes in the positions of its boundary atoms (i.e.,
atoms within two bonds from an atom on another processor). Also, some
atoms may move to another processor. To reduce the size of the boundary
region, we use three types of partitioning~\cite{stijnman}: standard cubic
(SC), which splits the simulation cell into $p=k^3$ subcubes; body-centred
cubic (BCC), which splits the cell into $p=2k^3$ truncated octahedra
centred at the lattice sites of the BCC lattice; and face-centred cubic
(FCC), which splits the cell into $p=4k^3$ rhombic dodecahedra centred at
the lattice sites of the FCC lattice. The new BCC and FCC partitionings
generate about 10\% less communication than the commonly used SC
partitioning. With these three partitionings, we can choose from a wide
range of processor numbers $p$, and in particular we can employ every
parallel computer where $p$ is a power of two.

We have implemented the parallel algorithm using the BSPlib communications
library~\cite{hill98} on a Cray T3E.  In the local relaxation, we have
achieved a speedup of 18.6 on 32 processors, corresponding to an
efficiency of 58\%. The efficiency loss is due both to the fluctuations in
work load per local iteration (caused by small differences in the numbers
of cluster atoms) and to waiting time at the end of successful time
intervals. At the optimum value of $b$, both effects are significant.
Losses due to the synchronizations themselves are negligible, as one
synchronization takes less than 1\% of the time of a local iteration. The
replication time was found to be the same as the time of 14 local
iterations, but overall this time is negligible since data only need to be
replicated after a successful attempt. In the global relaxation, we have
achieved a speedup of 19.3 on 32 processors for the 20,000-atom model,
corresponding to an efficiency of 60\%, see~\cite{stijnman} for more
details. Here, the efficiency loss is mainly due to redundant computations
for boundary regions, and communication of data for boundary atoms. The
total speedup of our parallel version depends on the mixture of local and
global relaxations needed. This mixture is influenced by a variety of
parameters such as the temperature $T$ in the Metropolis acceptance
criterion~(\ref{eq:metropolis}) and the expected reduction due to the
global relaxation. In our simulations, the amounts of CPU time spent on
local relaxation and on global relaxation were nearly equal.

\section{Results}

Using the scalable WWW algorithm we have generated one 10,000-atom
amorphous silicon network and one 20,000-atom network. In this section, we
discuss the structural and electronic properties of these networks. In
Table~\ref{tab:properties}, we compare our configurations relaxed with the
Keating potential with those of Djordjevi\'c, Thorpe, and
Wooten~\cite{djordjevic95} and with models generated by Barkema and
Mousseau using the improved WWW algorithm~\cite{barkemahq}. We also
provide irreducible ring statistics.

\begin{table*}
\caption{\label{tab:properties} Energetic and structural properties of
models relaxed with the Keating potential. The first two models, DTW4096a
and DTW4096b, are the 4096-atom models prepared
in~\protect\cite{djordjevic95} and refer, respectively, to a model with
and without four-membered rings.  Configurations BM1000a and BM1000b are
1000-atom configurations prepared by Barkema and Mousseau using the
improved WWW algorithm and BM4096 is a 4096-atom model prepared in the
same way~\cite{barkemahq}. Configurations `10k' and `20k' represent,
respectively, 10,000-atom and 20,000-atom models prepared using the
scalable WWW algorithm described in the text. The ring statistics are for
irreducible rings and $\rho_0$ is based on $d=2.35$\ \AA.}
\begin{tabular}{clllllll} \hline\hline
                         & DTW4096a & DTW4096b &BM1000a &BM1000b & BM4096 & 10k    & 20k    \\ \hline
$E$ (eV/atom)            & 0.336    & 0.367    & 0.267  & 0.264  & 0.304  & 0.301  & 0.286  \\
$\rho/\rho_0$            & 1.000    & 1.000    & 1.043  & 1.040  & 1.051  & 1.054  & 1.042  \\
$\langle r \rangle/d$    & 0.996    & 0.997    & 0.982  & 0.982  & 0.980  & 0.980  & 0.981  \\
$\langle \theta \rangle$ & 109.24 & 109.25 & 109.30 & 109.27 & 109.28 & 109.28 & 109.25 \\
$\Delta \theta$          & 10.51  & 11.02  & 9.21   & 9.20   & 9.89   & 9.88   & 9.63   \\
                         &        &        &        &        &        &        &        \\
rings/atom               &        &        &        &        &        &        &        \\
4                        & 0.015  & 0.000  & 0.000  & 0.000  & 0.000  & 0.000  & 0.020  \\
5                        & 0.491  & 0.523  & 0.472  & 0.480  & 0.490  & 0.480  & 0.456  \\
6                        & 0.698  & 0.676  & 0.761  & 0.750  & 0.739  & 0.742  & 0.759  \\
7                        & 0.484  & 0.462  & 0.507  & 0.515  & 0.467  & 0.512  & 0.501  \\
8                        & 0.156  & 0.164  & 0.125  & 0.116  & 0.148  & 0.142  & 0.149  \\
9                        &        &        & 0.034  & 0.033  & 0.035  & 0.034  & 0.039  \\ \hline\hline
\end{tabular} 
\end{table*}

Table~\ref{tab:properties} shows that the strain per atom for the
10,000-atom and 20,000-atom models is significantly lower than that of the
DTW models.  Compared to the 1000-atom models prepared with the improved
WWW algorithm (BM1000a and BM1000b) we find that the strain per atom in
our 10,000 and 20,000 atom models is only slightly higher, thus clearly
demonstrating the efficiency of the scalable WWW approach.

An important quantity that can be compared with experiment is the width of
the bond angle distribution $\Delta \theta$. Experimentally, this quantity
can be extracted from the radial distribution function
(RDF)~\cite{khalid99} or the Raman spectrum~\cite{beeman,vinkraman}. The
most recent measurement, obtained from the RDF, yields 10.45 degrees for
as-implanted samples and 9.63 degrees for annealed
samples~\cite{khalid99}. The bond angle distributions of the 10,000-atom
and 20,000-atom models generated by us are in good agreement with these
experimental values.

Although the Keating potential already produces high quality networks by
itself, it is important to check the stability of these networks when
relaxed with a more realistic interaction potential that does not require
a pre-set list of neighbors. For this purpose we use the Stillinger-Weber
(SW) potential~\cite{stillinger85} but with an enhanced angular force: the
three-body term is increased by 50\% with respect to the two-body term.
This ad-hoc modification was shown to produce good structural properties
for amorphous silicon~\cite{ding,holender91,art,art99,vinkmodsw}.

\begin{table*}
\caption{\label{tab:mswproperties} Structural properties of configurations
after relaxation with the modified Stillinger-Weber (mSW) potential. The
total ring number per atom (including reducible rings) is also reported,
as well as the energy after relaxation with the original Stillinger-Weber
(SW) potential.}
\begin{tabular}{cccccc} \hline\hline
                         &BM1000a &BM1000b & BM4096 & 10k    & 20k     \\ \hline
$E$ (eV/atom, mSW)       & -4.026 & -4.034 & -3.990 & -3.994 & -4.008  \\
$E$ (eV/atom, SW)        & -4.126 & -4.133 & -4.106 & -4.109 & -4.116  \\
$\rho/\rho_0$            & 0.947  & 0.950  & 0.936  & 0.938  & 0.933   \\
$\langle r \rangle/d$    & 1.018  & 1.017  & 1.020  & 1.021  & 1.020   \\
$\langle \theta \rangle$ & 109.25 & 109.24 & 109.20 & 109.19 & 109.20  \\
$\Delta \theta$          & 9.77   & 9.70   & 10.51  & 10.54  & 10.18   \\
                         &        &        &        &        &         \\
rings/atom               &        &        &        &        &         \\
4                        & 0.000  & 0.000  & 0.001  & 0.003  & 0.020   \\
5                        & 0.472  & 0.480  & 0.489  & 0.481  & 0.456   \\
6                        & 0.840  & 0.847  & 0.830  & 0.844  & 0.843   \\
7                        & 1.011  & 1.023  & 0.979  & 1.034  & 1.020   \\
8                        & 2.025  & 2.002  & 2.064  & 2.038  & 2.018   \\ \hline\hline
\end{tabular}
\end{table*}

The properties of the networks after relaxation with the (modified) SW
potential are reported in Table~\ref{tab:mswproperties}. For all
configurations, the bond angle distribution widens and the density
decreases.

\begin{figure}
\begin{center}
\includegraphics[width=8cm]{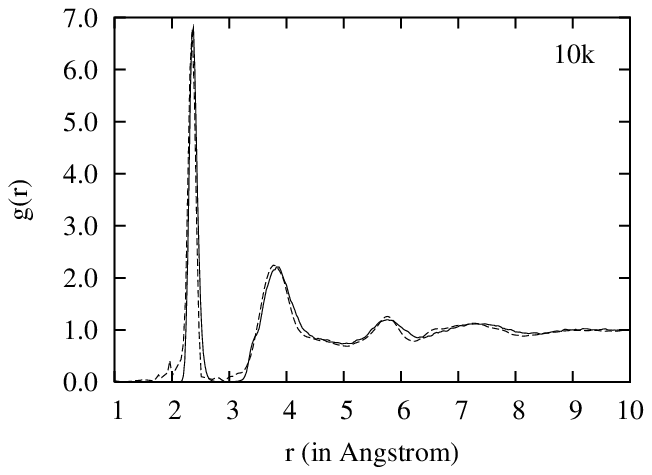}
\includegraphics[width=8cm]{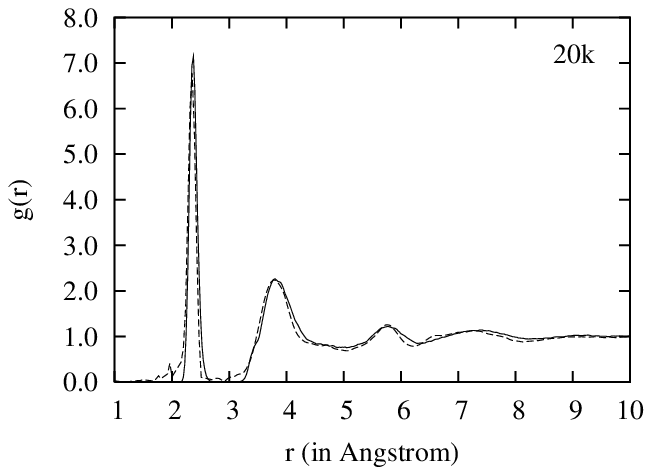}
\caption{\label{fig:rdf} Radial distribution function for the 10,000-atom
model (top) and the 20,000-atom model (bottom) after relaxation with the
modified Stillinger-Weber potential (solid line). The dashed line shows
the experimental result from Ref.~\onlinecite{khalid99}; distances are in
\AA.}
\end{center}
\end{figure}

Fig.~\ref{fig:rdf} shows the RDF for the 10,000-atom and 20,000-atom
models compared to the experimental RDF obtained by Laaziri {\it et al.}
on annealed {\it a}-Si samples prepared by ion
bombardment~\cite{khalid99}. Agreement is excellent. However,
configurations differing widely in topology can easily produce similar
RDFs. Agreement with the experimental RDF must therefore be regarded as a
minimum demand on a high quality CRN.

A more stringent criterion that can be used to evaluate the quality of a
model is the coordination number of the atoms. Using the minimum of the
RDF beween the first and second neighbor peak as the nearest neighbor
cut-off distance and after relaxation with the modified Stillinger-Weber
potential, we observe that the 10,000-atom and 20,000-atom model develop
0.9\% and 0.2\% of coordination defects, respectively.

\begin{figure}
\begin{center}
\includegraphics[width=8cm]{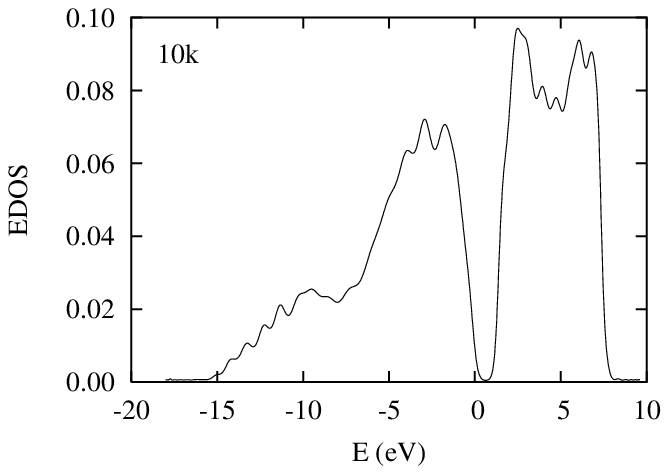}
\includegraphics[width=8cm]{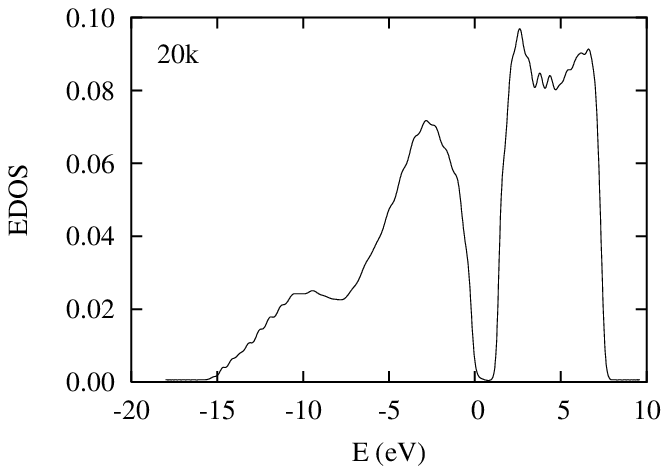}
\caption{Electronic density of states for the 10,000-atom model (top) and 
the 20,000-atom model (bottom) as obtained from ab-initio  
tight-binding~\cite{fireball}.}
\label{fig:edos}
\end{center}
\end{figure}

While structural averages provide good insight into the overall quality of
a model, they do not say much regarding local environments. It is
therefore also important to look at the electronic properties of our
models: even small densities of highly strained geometries or defects will
be picked up as states in the gap of the electronic density of states
(EDOS). In Fig.~\ref{fig:edos} we show the EDOS of the 10,000-atom and
20,000-atom models. The {\it Fireball} local-basis ab-initio
code~\cite{fireball} was used to obtain the EDOS. A remarkable feature of
the state densities shown here is the absence of states in the gap,
leading to a perfect gap of \mbox{1.3 eV} for both models.

\section{Conclusions}

We have presented here a scalable version of the WWW algorithm which
allows for local atomic rearrangements to be tried using only $\cal O$(1)
operations. We have developed an efficient parallel version which achieves
good load balance and limits communication. The scalable performance of
the algorithm has been demonstrated by generating one 10,000-atom and one
20,000-atom model. Structural and electronic properties of these models
are excellent and they compare well to experiments.

These high-quality models have the long term goal of accurately modeling
devices such as solar cells. At this point, using periodic-boundary
conditions in the two extended directions, we are able to simulate {\it
a}-Si films with thinkness of about \mbox{1000 \AA}. Once such atomic
configurations become available, the role of various structural and
electronic defects can be studied.

\section{Acknowledgements}   

We thank Dave Drabold for communicating to us the EDOS of the 10,000-atom
and 20,000-atom models.

\bibliographystyle{prsty}

\end{document}